\documentclass[final]{elsarticle}

\usepackage{lineno,hyperref}

\usepackage{amssymb}
\usepackage{enumerate}
\usepackage{multirow}
\usepackage{amsmath}

\usepackage{array}
\usepackage{algorithmic}
\usepackage{booktabs}
\usepackage{underscore}

\journal{Digital Signal Processing}

\begin{document}

\begin{frontmatter}

\title{Joint framework with deep feature distillation and adaptive focal loss for
weakly supervised audio tagging and acoustic event detection}

\author[a1]{Yunhao Liang}
\author[a1]{Yanhua Long \corref{cor1}}
\cortext[cor1]{Corresponding author}
\ead{yanhua@shnu.edu.cn}

\author[a2]{Yijie Li }
\author[a2]{Jiaen Liang }
\author[a3]{Yuping Wang }

\address[a1]{Key Innovation Group of Digital Humanities Resource and Research, \\
Shanghai Normal University, Shanghai, China}
\address[a2]{Unisound AI Technology Co., Ltd., Beijing, China}
\address[a3]{ByteDance, Beijing, China}

\begin{abstract}

A good joint training framework is very helpful to improve the performances
of weakly supervised audio tagging (AT) and acoustic event detection (AED)
simultaneously. In this study, we propose three methods to improve the
best teacher-student framework in the IEEE AASP Challenge
on Detection and Classification of Acoustic Scenes and Events (DCASE)
2019 Task 4 for both audio tagging and acoustic events detection tasks.
A frame-level target-events based deep feature distillation is first proposed, which
aims to leverage the potential of limited strong-labeled data in weakly
supervised framework to learn better intermediate feature maps.
Then, we propose an adaptive focal loss and two-stage training strategy
to enable an effective and more accurate model training, where
the contribution of hard and easy
acoustic events to the total cost function can be automatically adjusted.
Furthermore, an event-specific post processing is designed to improve the
prediction of target event time-stamps. Our experiments are performed
on the public DCASE 2019 Task 4 dataset, results show that
our approach achieves competitive performances in both AT (81.2\% F1-score)
and AED (49.8\% F1-score) tasks.

\end{abstract}


\end{frontmatter}


\section{Introduction}
\label{sec:intro}

Audio tagging (AT) refers to predict the category of acoustic events
that occur in an audio segment, while the task of acoustic event detection (AED)
not only needs to identify whether certain events occur in this segment,
but also needs to detect their onset and offset time-stamps.
The AED and AT technology can be applied in many areas, such as smart home \cite{debes2016monitoring},
health monitoring systems \cite{zigel2009method} and multimedia retrieval \cite{wold1996content, jin2012event}, etc.
Because labeling large-scale training data with detailed annotations is
high cost and time-consuming, the challenges of Detection and
Classification of Acoustic Scenes and Events (DCASE) from 2018 to
2020 Task 4 \cite{serizel2018large,turpault2019sound,turpault2020improving}
have been launched, to explore the possibility for exploiting a large
amount of unbalanced and unlabeled training data together with
a small weakly annotated training set to improve AED system performance.
These challenges have attracted increasing research attention in this
field \cite{serizel2019sound, shah2018closer, mcfee2018adaptive,huang2020guided}.
In this study, we also focus on this weakly supervised task to improve both of the
AT and AED system performances.

In the literature, many previous works have been proposed to improve the
AED systems from variety aspects, such as proposing new
back-end classifiers, new robust features, better post-processing methods,
and training data enhancement techniques, etc.
For the back-end classifiers, \cite{mcloughlin2015robust} proposed
to use a specially designed convolutional neural networks instead of the traditional
HMM and SVM to exploit a novel spectrogram image feature
for improving system performance.
In \cite{Hong2020}, authors proposed a gated multi-head attention pooling algorithm,
to attend the information of events from different heads at different positions.
Work in \cite{Kao2020} proposed a new classifier that consists of a modified DenseNet
as the feature extractor, and a global average pooling layer to predict the
frame-level labels at inference time. This classifier can directly
localize target events using the feature map
that extracted by DenseNet without any recurrent layers.
Authors in \cite{Park2020} proposed a two-stage polyphonic SED model
to handle the sound events overlapping in time-frequency.
A faster regional CNN and an attention-LSTM were used in the first stage
to capture those region-of-interests. In the second stage,
two CNNs combined with one LSTM were used for further
feature representation, the followed softmax combined with CTC was then
used for the final event detection and classification.
As mentioned above, most of those previously proposed back-end classifiers
focused on exploiting the discriminative information in different-level
feature representations.

At the feature level, most previous works focused on investigating
better robust feature learning techniques, such as
in \cite{Zhang2019,deBenito2020},
a Multi-Scale Time-Frequency Attention (MTFA) module
to enhance the acoustic features for AED was proposed.
It gathers information at multiple
resolutions to generate a time-frequency attention mask
which tells the model where to focus along both time and frequency axis.
With MTFA, the model could capture the characteristics of target events with different scales.
They achieved competitive results on DCASE 2017 Task 2.
In \cite{Lu2019}, authors focused on proposing new discriminative
feature learning method, they aimed to reduce the intra-class sample
distances and increase inter-class sample distances
rather than estimating the pair-wise distances of samples to
learn better class discrimination at feature-level. By
adding a distance metric constraint on feature extraction,
the learned features could have a good property for
improving the generalization of the classification models.
And works in \cite{Li2019} also focused on exploring
robust features for different sound classification tasks,
they introduced a multi-stream convolutional neural
network with temporal attention, to learn
a high-level robust feature representation from three
input streams, including the raw audio, the spectral features
and the temporal attention-based energy features.

During inference, post-processing is applied to smooth each event-class probability sequence.
In recent works, some better post-processing methods has been proposed. Such as,
in \cite{dinkel2020duration}, the impact of fixed-size window median filtering
was investigated, and they proposed a double thresholding as a more robust and
predictable post-processing method to improve the DCASE 2018 Task 4 AED performances.
And in \cite{miyazaki2020weakly}, three post-processing methods have been proposed
for smoothing the neural network outputs: the median filtering, accept gap, and
removal of the short duration events.

In addition, variety of data augmentation techniques are used in AED to improve
the generalization ability of models. For example,
\cite{li2019multi} perturbed the audio inputs by mixing in other environmental audio clips,
and leverage other training examples as sources of noise. \cite{lu2019semi} used
random time and frequency shifts as natural perturbations on audio data.
\cite{zhang2017mixup}
proposed a mixup augmentation to train the neural network
using convex combination of pairs of samples. In \cite{verma2019interpolation},
authors further extended this mixup to image detection tasks to provide a good consistency
perturbation for semi-supervised learning.

Looking at the recent approaches surveyed above, we see
all of the back-end classifiers, features, training data augmentation and
post-processing affect final AED and AT system performances.
However, for the weakly supervised AED tasks, the key issue is how to design a better
modeling framework to fully exploit the large-scale weakly and unlabeled
training data. Therefore, many recent works focus on the system modeling framework, semi-supervised or unsupervised learning approaches, such as, the joint training
system in \cite{joint1,joint2}. In DCASE 2018 Task 4, the baseline
used two convolutional recurrent neural network (CRNN) with a two-pass
training strategy to predict labels of unlabeled clips \cite{Serizel2018};
The best system of this challenge was based on the Mean-teacher (MT) model
\cite{tarvainen2017mean,jiakai2018mean}, in which two same CRNN networks
were taken as the teacher and student model with weight-averaged
consistency targets to exploit the large amount of unlabeled data;
In DCASE 2019 Task 4, the Guided Learning (GL) model \cite{lin2020guided}
convolutional system achieved the best results. The GL is also
a teacher-student (T-S) framework, but unlike the homologous T-S model
in MT, GL uses two different model structures to enable an appropriate
trade-off between AT and AED. These two models are trained synchronously and
forced to learn from the unlabeled data with tags generated by each other.

In this study, our work is also based on the Guided Learning architecture. The
contributions are: 1) we modify the GL to learn the models for AED and AT
as two independent branches in a joint framework; 2) a frame-level
target-events based deep feature distillation is proposed, and it aims to
leverage the maximum potential of limited strong-labeled data;
3) to enable an effective and more accurate model training,
a two-stage training strategy with an adaptive focal loss is further investigated;
4) an event-specific post processing is specially designed to
fix the prediction errors that result from outliers.
It is worth noting that, during the two-stage training,
the contribution of hard (difficult-to-classify) and easy (easy-to-classify)
acoustic events to the total cost function will be dynamically adjusted
in each iteration, which makes the importance of easy
events are down-weighted and the model can rapidly focus on hard events
in later stages. Experimental results on the DCASE 2019 Task 4 challenge show that
our proposed methods can achieve competitive performances in both AT
and AED tasks.

The rest of the work is organised as follows. In Section \ref{sec:proposed},
we introduce our proposed methods, including the AD and AED joint framework in
Section \ref{subsec:joint-fra}, the target-events based deep feature distillation
(TFD) in Section \ref{subsec:deep-fea}, the two-stage training
with adaptive focal loss in Section \ref{subsec:adapt-focal},
and the event-specific post processing (ESP) in Section \ref{subsec:post-process}.
Section \ref{sec:exp} presents the experiments, results and analysis in detail.
Finally, this work is concluded in Section \ref{sec:conclution}.

\section{Proposed method}
\label{sec:proposed}

In this section, we introduce our proposed methods on AED and AT tasks,
mainly focusing on the joint framework in Figure \ref{fig:model}. The
whole proposed framework is first introduced in section \ref{subsec:joint-fra}, followed
by the target-events based deep feature distillation (TFD) in section \ref{subsec:deep-fea},
the two-stage training with adaptive focal loss in section \ref{subsec:adapt-focal},
and in section \ref{subsec:post-process}, we present the details
of event-specific post processing (ESP).

\subsection{Joint framework}
\label{subsec:joint-fra}

\begin{figure*}[htbp]
  \scalebox{0.85}{\includegraphics[width=15cm]{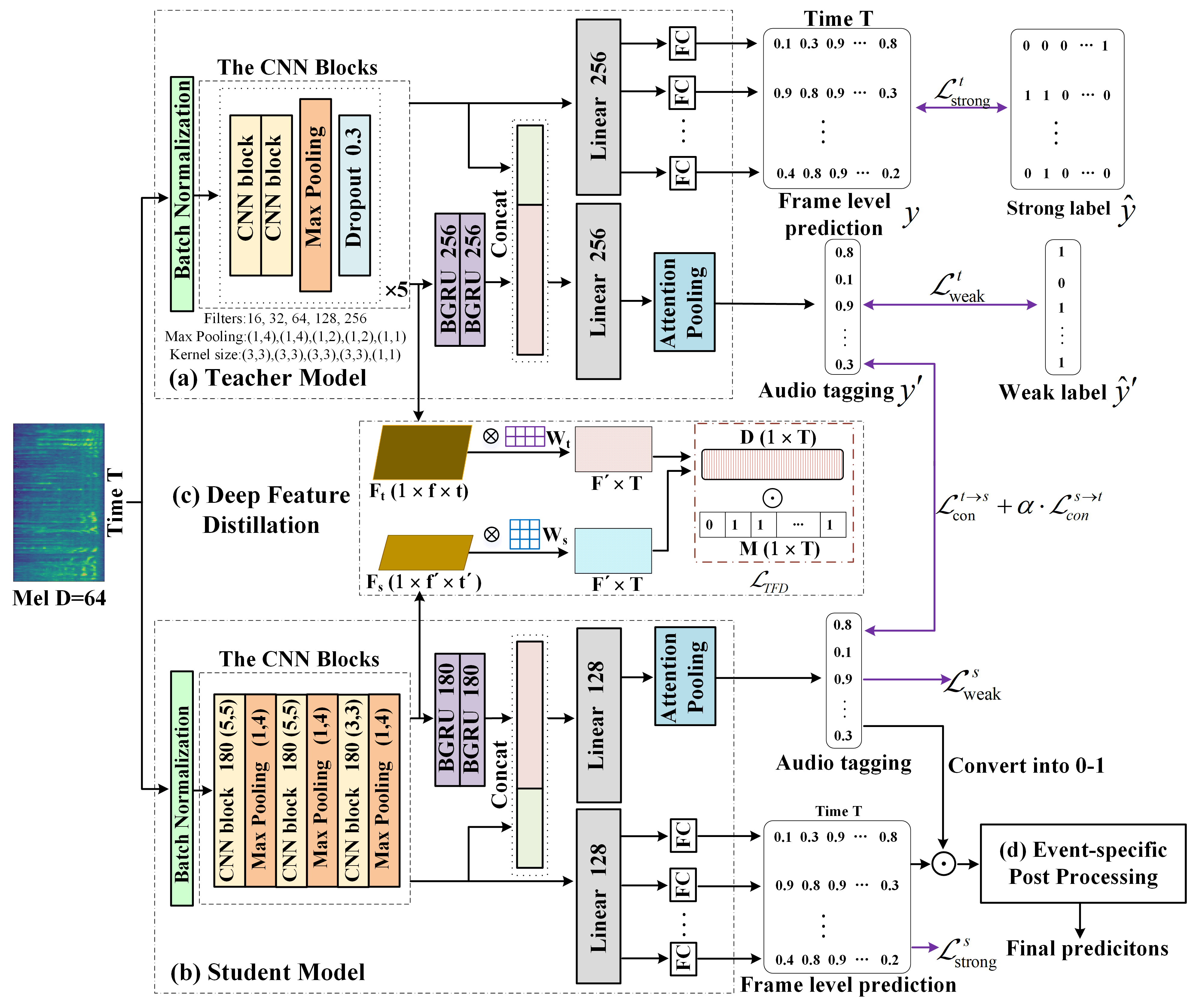}}
  \caption{Model architecture of the proposed joint framework for both audio tagging and acoustic event detection.}
  \label{fig:model}
\end{figure*}

\begin{figure}[htbp]
  \centering
  \includegraphics[width=6cm]{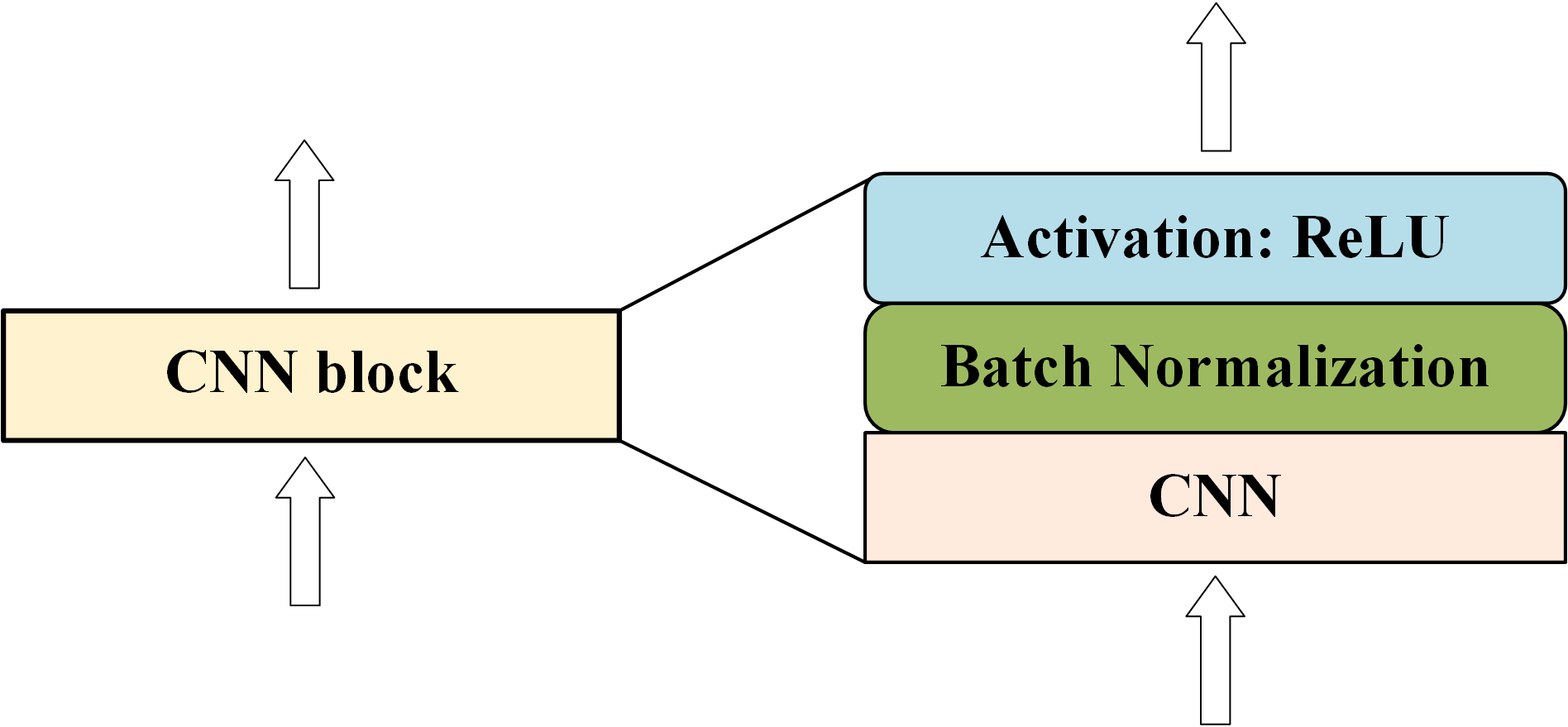}
  \caption{Structure of a single CNN block.}
  \label{fig:cnn}
\end{figure}

The proposed joint architecture for both weakly
supervised AT and AED is shown in Figure \ref{fig:model}.
It consists of four parts: the (a) teacher model, (b) student model,
(c) deep feature distillation module and (d) event-specific post processing module.
The teacher and student models are Convolution Recurrent Neural Networks (CRNNs),
but with different number of CNN blocks. The teacher model has five double-layer
CNN blocks with a larger time compression scale that professional for a better audio
tagging, while the student model only has three single-layer CNN blocks
with no temporal compression scale for a better event boundary detection.
Compared with the two same networks structure in conventional
Mean-teacher model \cite{tarvainen2017mean}, our smaller student model
can not only learn different feature information,
but also can reduce the model parameters and improve the training efficiency.
Figure \ref{fig:cnn} shows the detail structural of each single CNN block that included in
both teacher and student models. It is composed of a convolutional layer, batch normalization
and a ReLU activation function.

Compared with the Guided Learning model in \cite{lin2020guided} that is only composed of
two different convolution networks (CNN Blocks in (a) and (b)),
we propose to add two additional bi-directional Gated Recurrent Unit (BGRU) layers after the
CNN blocks for extracting the temporal information of CNN representations for a better audio tagging.
Different from GL and CRNN frameworks, we first
separate the AED and AT tasks into two independent branches as shown in
(a) and (b). The AED branch uses a fully connected layer with larger hidden states (the `Linear'
block) followed by ten small size separate fully connected layers (FC) with sigmoid activation
to model the outputs of CNN blocks for detection.
While in the AT branch, the outputs of both CNN Blocks and BGRUs are concatenated as the
input features of  `Linear' layer, followed by an attention pooling module \cite{jiakai2018mean}
with a consistency loss to perform the events classification. The motivation of this proposal
is that, the design of AED system structure requires to accurately reflect the frame-level
time-stamps information for detection, and it trains on strongly labeled data, while the AT system does not. Thus,
by sharing the same standard feature encoder of CNN Blocks, separate
AT and AED into two independent branches can provide
possibility for improving the two different tasks independently, such as the
two additional BGRU layers designed for the AT branch.  The
temporal feature representations produced by BGRUs
can capture robust sequential information in the audio, which can
provide significant complementary information to the standard feature
representations from CNN Blocks.

Besides the independent AED and AT branch, we also propose
a deep feature distillation in module (c), where only the
frame-level feature representations of target-events are used to compute the distillation
loss, which can maximally leverage the information of limited strongly
labeled data to regulate the teacher-student model training.
Moreover, a two-stage model training strategy with adaptive focal loss
is investigated, it can automatically exploit the importance of hard training
acoustic events and achieve more stable models. Finally, an
event-specific post processing (module (d)) is proposed to further improve
the accuracy of final predictions.
In addition, as shown in the block between student model and event-specific
post processing, we apply an element-wise multiplication between the one-hot
audio tagging prediction labels and the frame-level detection posteriors
to ensure the consistency of detection and audio classification results.

\subsection{Target-events based deep feature distillation (TFD)}
\label{subsec:deep-fea}

Knowledge distillation (KD) is one popular technique that used in machine learning
to improve a smaller student model given large teacher model guiding \cite{huang2020multi, shi2019teacher, romero2014fitnets}.
It can be performed not only on the output logits \cite{hinton2015distilling, pilzer2019refine} between teacher and student model,
but also on activations, neurons, or intermediate features \cite{cheng2020explaining, mirzadeh2019improved, dong2017improving}. For the feature-based
knowledge distillation, most related works choose to distill
the intermediate information between teacher and student model
using the whole feature maps directly, including the intra-utterance
similarity preserving knowledge distillation that  proposed in \cite{chang2020intra}
for audio tagging, the similarity matrix is also calculated on the whole feature maps.

In this study, our goal is to improve the target-events boundary
detection performance in weakly supervised AED task,
how to maximally exploit the frame-level alignment information
of the available limited strong-labeled data is crucial. Therefore,
instead of using the whole feature maps to
perform the knowledge distillation, we propose to force the distillation
performed only on the frame-level deep feature vectors that belong to target-events.
The specific operations are illustrated in module (c) of Figure \ref{fig:model}
and we define the distillation loss function as,
\begin{equation}
  \mathcal{L}_{TFD} = \frac{1}{T} \left( \left( \mathbf{D} \odot \mathbf{M} \right) \cdot \mathbf{A} \right)
  \label{eql-simi}
\end{equation}
where $\mathbf{M} \in \mathbb{R}^{1 \times T}$ is the frame-level label mask
vector with $j$-th element $\mathbf{M}_{j} \in [0,1]$, and 1 means $j$-th frame
is labeled as one type of target events in the limited strong-labeled dataset.
$\mathbf{A} = [1,1,...,1]_{T \times 1}^\top$, $T$ is the frame size, $\odot$ and
$\cdot$ denote the element-wise and normal matrix multiplication respectively.
$\mathbf{D}\in \mathbb{R}^{1 \times T}$ is a frame-level similarity matrix,
where  $\mathbf{D}_{j}, j = 1, 2, ..., T $ represents $j$-th frame Euclidean distance between
the transformed deep feature vectors of CNN blocks outputs in teacher and student model, and it
is defined as,
\begin{equation}
  \mathbf{D}_{j} = {\parallel (\mathbf{F}_{t} \cdot \mathbf{W}_{t})_{j} - (\mathbf{F}_{s} \cdot \mathbf{W}_{s})_{j} \parallel}_{2}
  \label{eql-dis-pt-ps}
\end{equation}
where $(\mathbf{F}_{t} \cdot \mathbf{W}_{t})_{j}$ and $(\mathbf{F}_{s} \cdot \mathbf{W}_{s})_{j}$
is $j$-th column of transformed
feature maps $\mathbf{F}_{t}$ and $\mathbf{F}_{s}$ respectively,
$\mathbf{W}_{t}$ and $\mathbf{W}_{s}$ are their transformation matrix
as shown in module (c) of Figure \ref{fig:model}.

By introducing the distillation loss $\mathcal{L}_{TFD}$
into the total training loss of  weakly supervised framework,
we hope it can maximally leverage the information of limited strong-labeled data
to regulate the teacher-student model training, especially for
guiding the early training stages of T-S model, because
all the information captured by the $\mathcal{L}_{TFD}$ is derived
from those target-events with golden-standard time-stamps.

Different from the proposed TFD, similar to \cite{chang2020intra},
most conventional deep feature distillation works
perform the deep feature KD directly on the whole transformed maps
$(\mathbf{F}_{t} \cdot \mathbf{W}_{t})$, $(\mathbf{F}_{s} \cdot \mathbf{W}_{s})$
using the conventional Mean Squared Error (MSE) as similarity measure (DFD-MSE),
or the frame-level Euclidean distance similarity matrix $D$ (DFD-D).
The advantage of DFD-MSE and DFD-D is that, they can use all the available
strongly-labeled, weakly and unlabeled training data for the knowledge
distillation, instead of the limited strongly-labeled data in our proposed
TFD. Experimental comparisons are presented in Section \ref{sec:result}.

\subsection{Two-stage training with adaptive focal loss}
\label{subsec:adapt-focal}

Under the weakly supervised AED task like DCASE 2019 Task 4, three
types of dataset with different level of annotations are  provided,
i.e., a limited strong labeled set with time-stamps
for target acoustic events, a small weakly annotated set
with only the multiple events presence labels (without time-stamps),
and a large amount of unlabeled training dataset. To make full use of
all the available datasets, recent work in \cite{lin2020guided}
proposed a Guided Learning (GL) strategy, in which
the total connectional binary cross entropy (BCE) loss of the
weakly supervised T-S training consists of
two parts as follows:
\begin{equation}
\label{Loss'}
{\cal L} = {\cal L}_{weak}^{t,s} + {\cal L}_{strong}^{t,s} + {\cal L}_{con}^{t \to s} + \alpha  \cdot {\cal L}_{con}^{s \to t}
\end{equation}
where ${\cal L}_{weak}^{t,s}={\cal L}_{weak}^{t}+{\cal L}_{weak}^{s}$,
${\cal L}_{strong}^{t,s}={\cal L}_{strong}^{t}+{\cal L}_{strong}^{s}$
are the clip-level and frame-level supervised loss as shown in Figure \ref{fig:model} for the AT
and AED respectively. The last two terms are the clip-level consistency
losses performed on all types of training dataset.
${\cal L}_{con}^{t \to s}$ denotes using teacher predictions to
guide the student model training, while ${\cal L}_{con}^{s \to t}$ denotes
using the student to fine-tune the teacher model with a small weight $\alpha$.
During the earlier T-S training, only the first three terms are used,
the last term is normally added when the teacher model becomes
relatively stable. More details can be found in \cite{tarvainen2017mean, lin2020guided}.
In Equation (3) and the following Equation (6), all the loss with `weak' subscript are computed using
the weakly labeled training data, while the loss with `strong' and `con' subscripts
are computed using the strongly labeled and all types of training data,
including the strongly, weakly and unlabeled parts, respectively.

\begin{figure*}[!htbp]
  \centering
  \includegraphics[width=9cm]{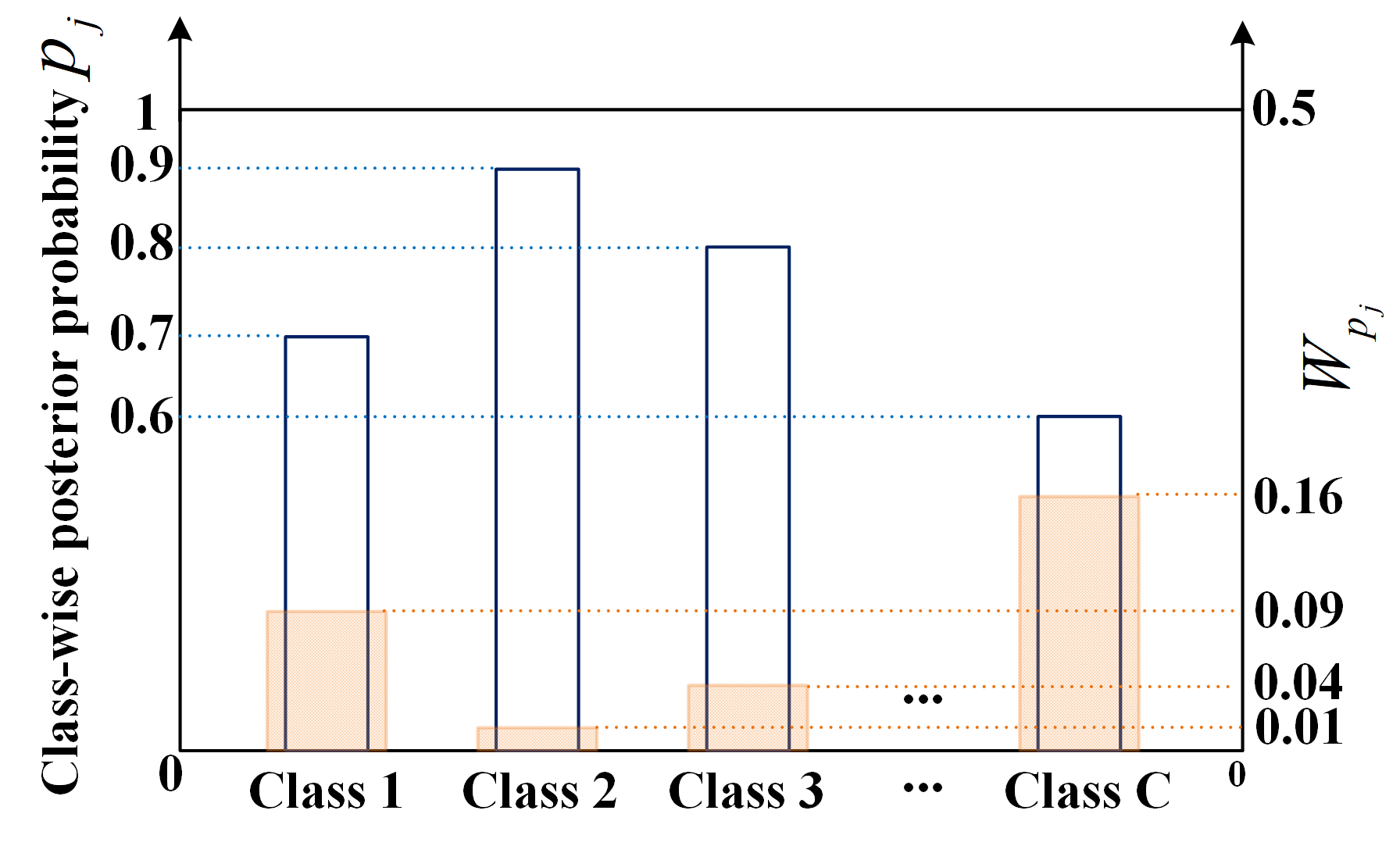}
  \caption{Illustration of class-wise adaptive weights distribution in ${\cal L}_{af}$.}
  \label{fig:adpatfl}
\end{figure*}

Motivated by the principle of focal loss in \cite{lin2017focal}, here
we aim to improve the GL training by combining the above BCE loss
with an adaptive focal loss that is defined as follows:
\begin{equation}
\label{Laf}
{{\cal L}_{af}} =  - \frac{1}{CK}\sum\limits_{j = 1}^C{\sum\limits_{i = 1}^K {\left( {1 - p_{ {ij} }^\gamma } \right)} }  \cdot \log \left( {{p_{ {ij} }}} \right)
\end{equation}
where $\gamma$ is a scaling factor to control the loss
contribution of posterior probability $p_{ij}$ for $i$-th clip, $j$-th target-event
category. $K$ is the total size of audio clips with both weakly and
strong labels in a minbatch, $C$ is the number of target-event
categories. Figure \ref{fig:adpatfl} illustrates the distribution of class-wise weight
in the ${\cal L}_{af}$ that varies with its corresponding posterior probability at one training
epoch. In Figure \ref{fig:adpatfl}, the left Y-axis represents the
class-wise posterior probability $p_{j} = \sum\limits_{i = 1}^K p_{ij}$,
and the right Y-axis represents the corresponding
class-wise adaptive weights $W_{p_{j}} = \sum\limits_{i = 1}^K \left({1 - p_{ij}}^\gamma\right)$.
From Figure \ref{fig:adpatfl}, it's clear to see that the ${{\cal L}_{af}}$
tends to assign a higher weight to the hard (small $p_{ij}$)
events than the easy ones, it automatically forces the model pays more
attention to the learning of this hard category in next training
iteration, and it is dynamically adapted in each training epoch.

Moreover, to guarantee the basic ability of BCE loss for achieving
a relatively stable T-S model, we find that the adaptive focal
loss should be utilized in a curriculum learning \cite{bengio2009curriculum}
way. Therefore, we propose a two-stage training strategy to train the whole
teacher-student model of Figure \ref{fig:model} as follows:

\verb"Stage 1", performing the normal GL training together with the proposed
TFD in first $\tau$ epochs using:
\begin{equation}\label{L-s1}
{\cal L}_{s1} = {\cal L} + \beta  \cdot {{\cal L}_{TFD}}
\end{equation}
where $\beta \in [0,1]$ is used to control the distillation depth.

\verb"Stage 2", performing the fine-tuning epochs by replacing the
${\cal L}_{weak}^{t,s}$, ${\cal L}_{strong}^{t,s}$ in ${\cal L}$ using their
corresponding adaptive focal loss ${\cal L}_{af - weak}^{t,s}$, ${\cal L}_{af - strong}^{t,s}$
respectively as,
\begin{equation}
\label{ls2}
\begin{split}
{\cal L}_{s2}  & = {\cal L}' + \beta  \cdot {{\cal L}_{TFD}} \\
{\cal L}'  = {\cal L}_{af - weak}^{t,s} & + {\cal L}_{af - strong}^{t,s} + {\cal L}_{con}^{t \to s} + \alpha  \cdot {\cal L}_{con}^{s \to t}
\end{split}
\end{equation}

By performing the adaptive focal loss together with two-stage
weakly supervised training, we hope that the class imbalance issue in training data,
and various difficulty level of AED between multiple target events
will be dynamically regulated and focused during model training.
Because in \verb"Stage 1", the discriminations between those easy-to-classify
acoustic events are well-learned with ${\cal L}_{s1}$, and it results in
relatively stable teacher and student models. As the training progresses
in \verb"Stage 2", the adaptive focal loss
can automatically down-weight the contribution of easy events during training,
and this makes the training rapidly focus the model on difficult-to-classify events.
By using the two-stage training, the ${\cal L}_{TFD}$, ${\cal L}_{af}$ and the BCE
losses are well integrated to boost the whole weakly supervised system performances.

\subsection{Event-specific post processing (ESP)}
\label{subsec:post-process}

In both audio tagging and acoustic event detection tasks, the
frame-level prediction outputs of the model may be non-consecutive, such as,
many detection outliers may produce too many extremely short-duration
target events occurrence, resulting in inaccurate time-stamps.
Therefore, the traditional way is to apply linear or non-linear
filters to smooth the prediction outputs. In DCASE Challenges,
Median filtering (MF) has proven to be an very effective method
in smoothing the noisy outputs of the student model for AED tasks,
such as in \cite{turpault2019sound}, authors used median filters
with the same window size for any sound event class detection.

However, DCASE 2019 Task 4 is a multiple target events detection instead of
single one, the subsequent duration of each event in audio clip
varies significantly. Conventional median filtering with
fixed window size is no longer suitable for this task. Recent
works in \cite{delphin2019mean,lin2020guided} used group of median
filters with adaptive window size by calculating the average
duration of events with strong labels on the development set.
However, each event duration is not an uniform distribution, using the average
duration to optimize the median filtering window size may not the optimal.
So, we propose to use event-specific MF window size as:

\begin{equation}
   {\cal W}_{c} = \left(\frac{1}{N_{c}}\sum\limits_{i = 1}^{N_{c}} L_{i} \right) \cdot \eta
  \label{eql-win}
\end{equation}
where ${\cal W}_{c}, c=1, 2, ..., C$ is the MF window size of
class $c$, $N_{c}$ is the segment index for the inflection point of cumulative
distribution of short-to-long sorted segments of the $c$th class target event.
$L_{i}$ is the duration of $i$-th segment of event $c$.
$\eta$ is a scaling factor and set to $1/3$ in our experiments.
All strong labeled training clips are used to compute ${\cal W}_{c}$.

\begin{figure*}[!htbp]
  \centering
  \includegraphics[width=12cm]{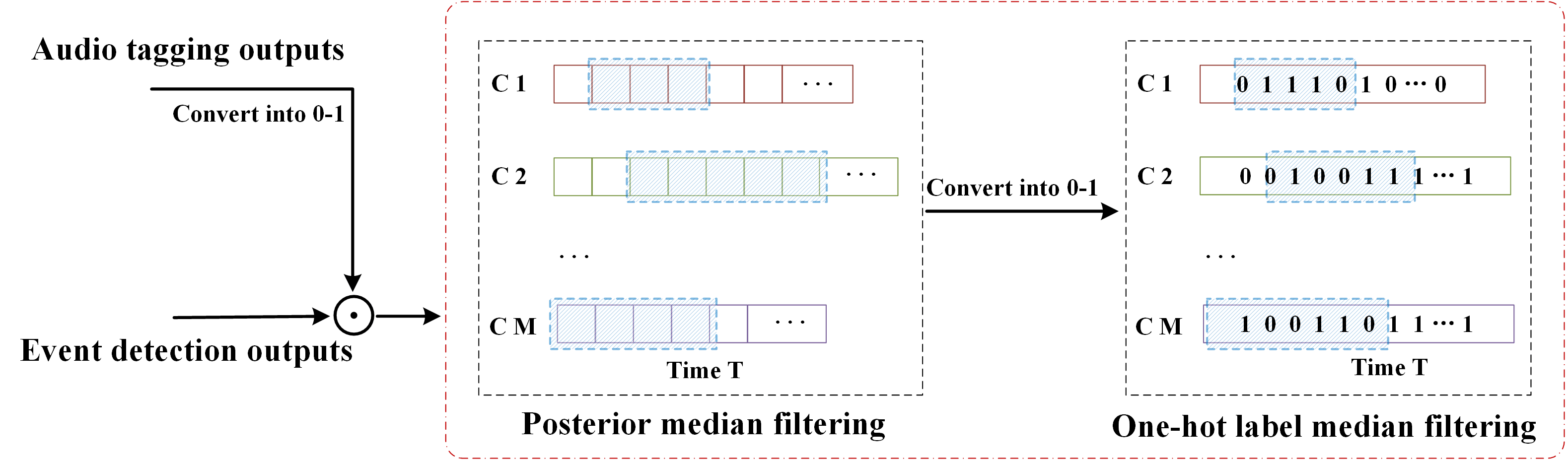}
  \caption{Event-specific post processing.}
  \label{fig:esp}
\end{figure*}

As shown in Figure \ref{fig:esp}, we perform the proposed event-specific
post processing with Equation (\ref{eql-win}) on both
the frame-level posteriors and one-hot predictions using the same
${\cal W}_{c}$. As shown in the bottom right-hand corner of Figure \ref{fig:model},
instead of feeding the frame-level acoustic
event detection outputs into the proposed ESP directly, here, in Figure \ref{fig:esp},
we propose to perform an element-wise
multiplication between the one-hot audio tagging prediction labels
and the frame-level detection posteriors. This operation can not only
ensure the consistency of acoustic event detection and audio tagging
results, but also can improve the AED performances by integrating the
audio tagging guidance.

\section{Experiments and Results}
\label{sec:exp}

In this section, we first describe the dataset of
DCASE 2019 Task 4 challenge in Section \ref{subsec:data}, then,
all the experimental system configurations are presented in
Section \ref{subsec:cfg}, followed by the results and discussions
in Section \ref{sec:result}.

\subsection{Dataset}
\label{subsec:data}

Our experiments are performed on the dataset of DCASE 2019 Task 4 Challenge \cite{turpault2019sound} \footnote{http://dcase.community/challenge2019/task-sound-event-detection-in-domestic-environments}. It is a sound event detection task in domestic environments.
It aims to provide not only the event class (AT task), but also the event time boundaries given that multiple events can be present in an audio recording (AED task).
The challenge of this task is to explore the possibility to exploit a large amount of unbalanced and unlabeled training data together with a small weakly annotated training set to improve system performance remains but an additional training set with strongly annotated synthetic data is provided.

The dataset for this task is composed of 10 sec audio clips recorded in domestic environment or synthesized to simulate a domestic environment. The focused 10 class sound events are: speech, dog, cat,
alarm/bell/ringing, dishes, frying, blender, running water, vacuum cleaner and the electric shaver/toothbrush. The training set includes 1,473 weakly labeled,
13,390 unlabeled and 2,045 strongly labeled audio clips.
1,077 and 692 strongly labeled clips are taken as the development
and evaluation clips respectively.

\subsection{System configurations}
\label{subsec:cfg}

We extract 64 log mel-band magnitudes as features,
each 10-second clip is transformed into 600 frames.
Details of our T-S model structure and its parameters for each block
is shown in Figure \ref{fig:model} and
section \ref{sec:proposed}.
We take both the MT \cite{jiakai2018mean} and GL \cite{lin2020guided}
as baselines, because the MT was the official
baseline, while GL was the best single system for DCASE 2019 Task 4 and our
proposal is improving the GL system.
$\alpha  = 1 - {\lambda ^{e - s}}$, $\beta  = {e^{ - 5{{\left( {1 - x} \right)}^2}}}$,
where $e$ is the current epoch and $x \in \left[ {0,1} \right]$.
$\alpha$ is a minimum value, it controls the student model guidance
to fine-tune the weight of the teacher model and the consistency cost
coefficient $\beta$ was ramped up from 0 to their maximum values,
using a sigmoid-shaped function.
The $\alpha$ and $\beta$ are gradually changing with the model training.
More details can reference \cite{tarvainen2017mean}.
The $\gamma$, $\lambda$ and $s$ is set to 2, 0.996 and 30 respectively in the two-stage training.

$F1$-score is used to measure the system performances.
Not as the traditional recall and precision rates, the $F1$-score
takes into account both the precision and recall statistics
of the classification and detection model, and it can be
regarded as a harmonic average of the precision and recall.
Using $F1$-score can effectively avoid the impact on the
analysis of model performance when the two indicators
of precision and recall conflict, and can evaluate the
system performance more accurately and conveniently.
The acoustic event detection uses an
event-based $F1$ (event-F1) that computed
with a 200 ms collar on onsets and a 200 ms / 20\% of the
events length collar on offsets, and a segment-based $F1$ (segment-F1)
on 1 second segment is taken as a secondary measure.
Segment-based metrics compare system output and reference in short time segments.
Active/inactive state for each event class is determined in a fixed length
interval that represents a segment. This alleviates issues related to annotator
subjectivity in marking onset and offset of sound events. Event-based metrics
compare system output and the corresponding reference event by event.
The overall (macro-average) $F1$-score (AT-F1) at clip
level is used to measure the audio tagging performances.
The metrics are computed using the sed_eval library \cite{mesaros2016metrics}.

\subsection{Results and Discussion}
\label{sec:result}

This subsection presents the results and discussions
of our proposed framework, including the overall results
of all the proposed techniques in section \ref{subsec:overall},
the analysis of system training parameters tuning of
the adaptive focal loss in section \ref{subsubafl}, and followed
by the statistical analysis, and class-wise performance comparison
of the event-specific post processing.

\subsubsection{Overall validation of the proposed methods}
\label{subsec:overall}

All techniques proposed in this study are examined on DCASE 2019 Task 4
evaluation set. Results are shown in Table \ref{tab:result-all}.
The `MT' and `GL' are the official baseline and top-1 ranked submission single
system in this challenge, respectively, we take both of them as our system
performance comparison baseline systems.
`PTS' is the backbone T-S model of our proposed framework in Figure \ref{fig:model},
without the proposed target-events based deep feature distillation (TFD),
two-stage training with adaptive focal loss (AFL) and
the event-specific post processing (ESP) techniques.
Technical improvements of PTS over GL is presented in section \ref{subsec:joint-fra}.
`DFD-MSE' and `DFD-D' represent performing the deep feature
KD on two transformed feature maps $\mathbf{F}_{t} \cdot \mathbf{W}_{t}$,
$\mathbf{F}_{s} \cdot \mathbf{W}_{s}$ using the conventional MSE criterion \cite{chang2020intra}, and on the frame-level
Euclidean distance similarity matrix $D$ respectively.
Both of them are performed on all the training data because
they do not need any strong-label information. `DFD-D(w+u)' means
only performing the `DFD-D' on the weakly and unlabeled training clips.

\begin{table}[!ht]
\caption{Development/Evaluation set $F1$-scores (\%) of the proposed methods, the
DCASE 2019 Task 4 official baseline (MT) and top-1 ranked single system (GL).}
\label{tab:result-all}
\centering
\begin{tabular}{llccc}
\toprule
ID & System & Event-F1 & Segment-F1 & AT-F1   \\
\midrule
0 & MT (Official baseline \cite{jiakai2018mean})  & -/~25.8        & -/~53.7          & -/~45.8 \\
1 & GL (Top-1 \cite{lin2020guided})    & \textbf{43.3}~/~\textbf{40.5}        &\textbf{67.1}~/~\textbf{ 66.5}          & \textbf{74.1}~/~\textbf{70.2} \\
\midrule
2 & PTS(w/o BGRUs)                 & 42.0~/~39.4        & 63.3~/~61.0          & 71.2~/~69.1 \\
3 & PTS                 & 43.1~/~40.9        & 69.0~/~68.1          & 75.0~/~73.4 \\
4 & PTS+DFD-MSE                       & 44.1~/~41.3        & 69.9~/~69.6          & 71.3~/~68.7 \\
5 & PTS+DFD-D                 & 44.4~/~41.7        & 73.2~/~71.1          & 73.4~/~72.9 \\
\midrule
6 & PTS+TFD                     & 48.0~/~45.4        & 71.7~/~70.2          & 78.6~/~77.1 \\
7 & PTS+TFD+DFD-D(w+u)            & 42.9~/~40.1        & 72.9~/~72.0          & 75.2~/~73.3 \\
8 & PTS+TFD+AFL                  & 48.9~/~47.1        & 75.6~/~74.0          & 80.0~/~78.1 \\
9 & PTS+TFD+AFL+ESP              & \textbf{51.6}~/~\textbf{49.8}        & \textbf{76.4}~/~\textbf{75.9}          & \textbf{83.4}~/~\textbf{81.2} \\
\bottomrule
\end{tabular}
\end{table}

In Table \ref{tab:result-all}, by comparing system 0 and 1, it is clear
that, the best single system GL in DCASE 2019 Task 4 challenge
outperforms the official baseline MT model significantly.
When we compare system 2 with 1, it is clear that, the proposed backbone T-S network
without BGRUs feature concatenation is much worse than GL system.
The independent branch design of AED and AT shows no advantage.
However, when we combine the feature representations from BGRUs
with the standard deep features from CNN Blocks, the performances of
proposed T-S model significantly improved, it achieves
0.4\%, 1.6\% AED and 3.2\% AT $F1$-score improvements over GL on the real
evaluation set. It indicates that, after being combined
with the independent design of detection and classification
branches, the advantage of complementary information in
temporal sequential features from BGRUs are well exploited.

Moreover, when comparing system 6, 8 and 9 with system 3,
continuous performance gains are obtained by
adding the proposed TFD, AFL and ESP techniques into the
backbone model, either on the development set or the real evaluation test set.
Such as, comparing system 6 with 3, the target-event based
deep feature distillation improves the event-F1 from
43.1\% to 48.0\% on development set, and 41.3\% to 45.4\%
on evaluation set, the AT-F1 also has more than 3.0\% absolute
gains on both test sets. Although the segment-F1 improvements
are relatively slight, these performance gains are good to
show the effectiveness of the proposed TFD, because
as presented in Section 2.2, the principle of TFD is to maximally
utilize the strongly labeled target-events frame-level boundary
information.

\begin{figure*}[!ht]
  \centering
  \includegraphics[width=12cm]{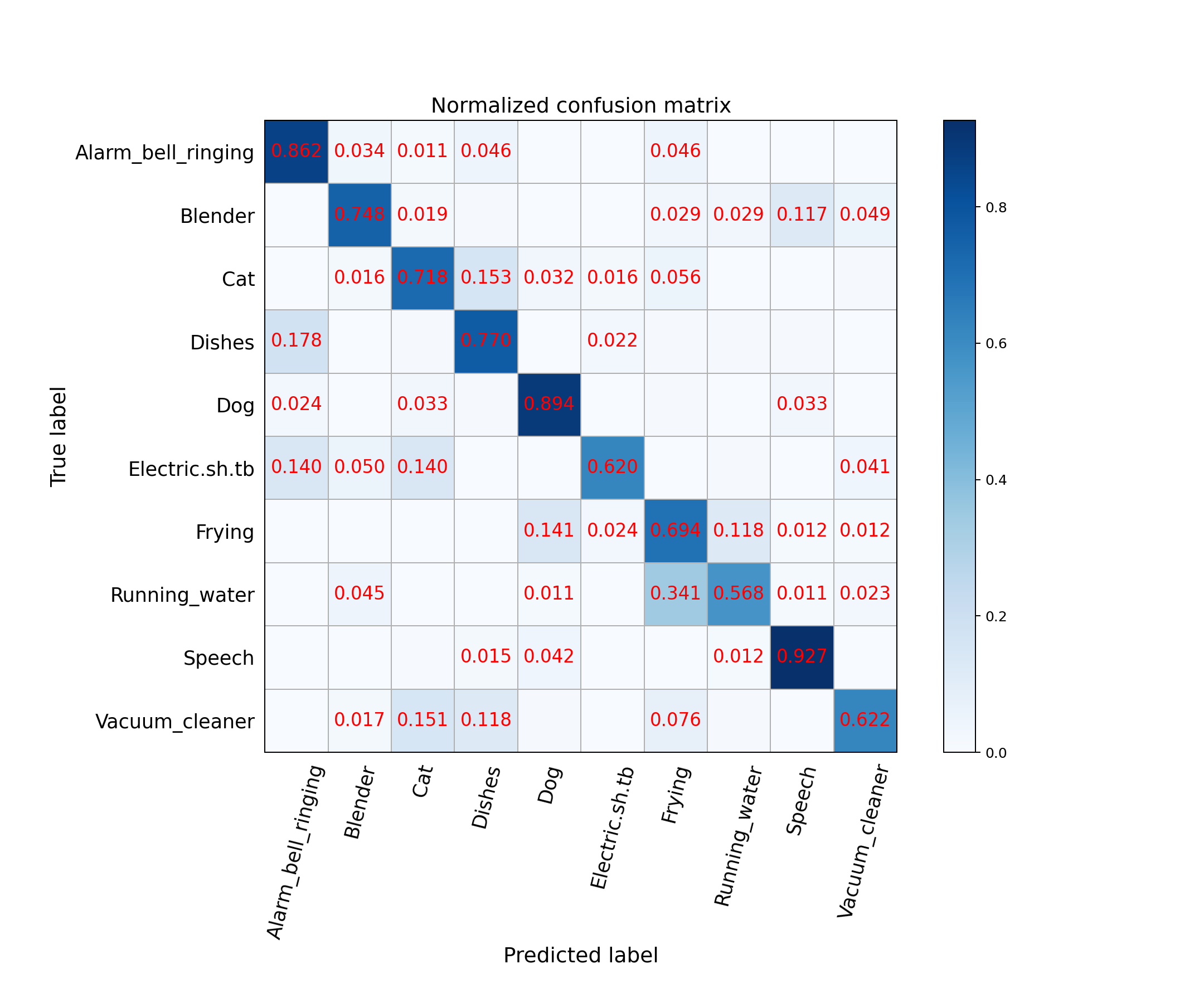}
  \caption{Confusion matrix heat map image of system 2 (the baseline) on the
  evaluation set for audio tagging.}
  \label{fig:confusiongl}
\end{figure*}

\begin{figure*}[!ht]
  \centering
  \includegraphics[width=12cm]{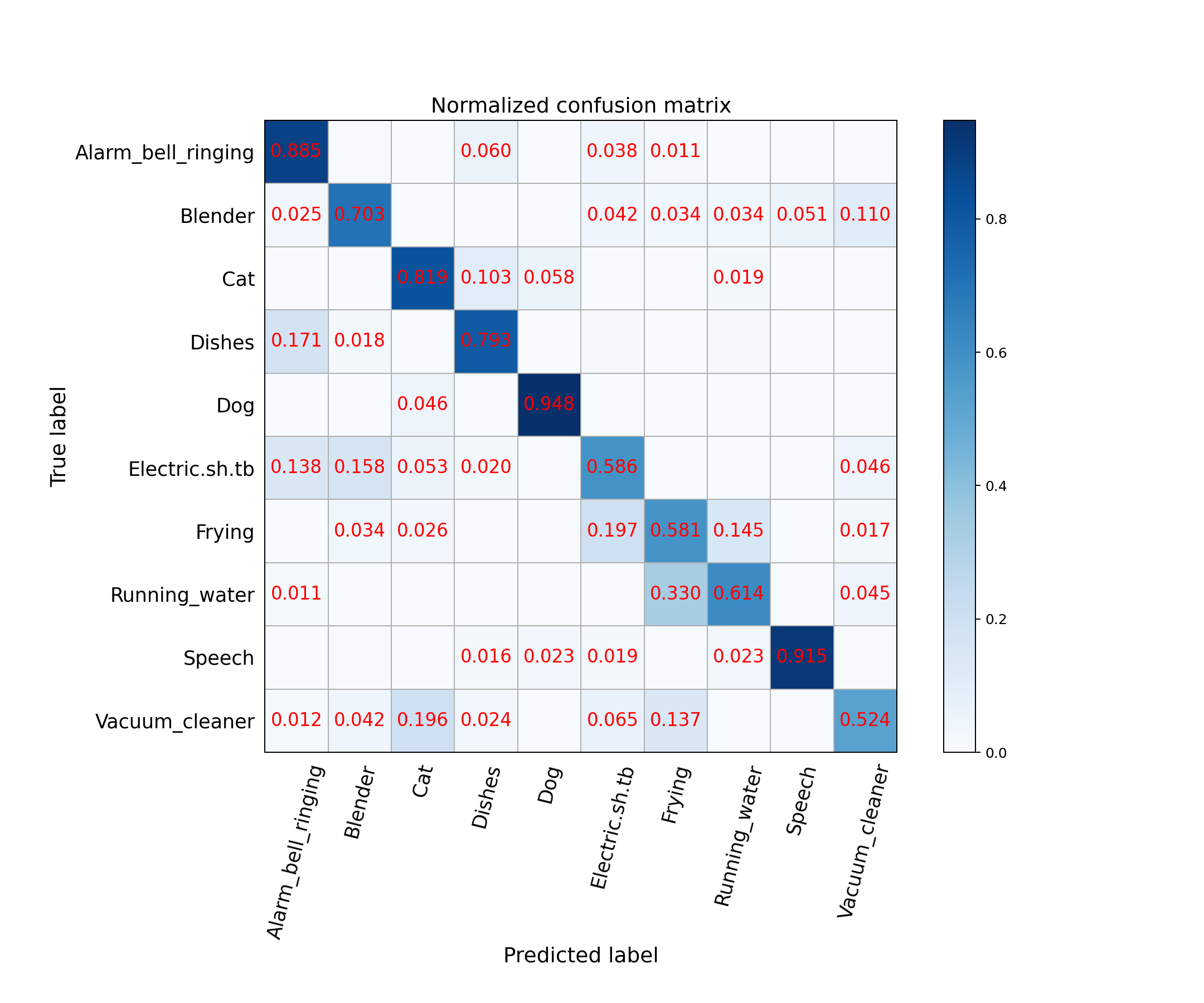}
  \caption{Confusion matrix heat map image of system 9 (the best) on the
  evaluation set for audio tagging.}
  \label{fig:confusion}
\end{figure*}

Comparing system 8 with 6, we see absolute 0.9\% to 3.9\%
F1 performance gains are achieved on both development and evaluation sets.
It means that the proposed two-stage training with adaptive
focal loss outperforms the one-stage training strategy significantly,
because all the systems from 2 to 7 are trained only using the standard
one-stage teacher-student BCE loss as in the GL system. By introducing the
adaptive focal loss, we change the system training from standard one-stage to
the two-stage way. The consistent performance gains on both
development and evaluation set reflect that,
the proposed AFL not only enhances the system performances,
but also provide more robust and stable acoustic models.

Finally, in system 9, we achieve the best results for both the
acoustic event detection and audio tagging tasks. Such as,
on the evaluation set, this system outperforms the GL significantly by
absolute 9.3\%, 9.4\% and 11.0\% $F1$-scores
in event-based, segment-based AED and macro-average AT-F1, respectively.
We think that these improvements are good enough to prove the effectiveness
of our proposed techniques. In addition,
when comparing system 9 with 8, 0.8\% to 3.4\% absolute F1 improvements
are obtained, especially for the AT task ($>3.0\%$ gain), it indicates
that, the event-specific post processing can smooth the short-duration
outliers very well. Furthermore, to show more detail classification performances
of individual acoustic classes, in Figure \ref{fig:confusiongl} and Figure \ref{fig:confusion},
we present the classification confusion matrix of system 9 (the best)
and system 2 (baseline system)
for AT on the evaluation set as a heat map image.
By comparing these two images, it's clear that the AT performance of
almost all the events are significantly improved.
Moreover, in Figure \ref{fig:confusion},
we note that the dog, speech and alarm bell ringing are much easier to
be predicted than other acoustic events; The most
difficult classes to predict are the electric shaver toothbrush,
frying, and vacuum cleaner, almost 50\% of them are misclassified as other events.

In addition, to further validate the proposed TFD, system 4, 5 are built for
performance comparison, we see that there are big performance gaps between
system 4, 5 and system 6, it tells us that the conventional deep feature
distillations performed on the whole feature maps are not better
than the target-event based one, even they can be trained using all
the training data. By comparing system 5 with 4,
we see that performing knowledge distillation using frame-level
Euclidean similarity is better than using the conventional MSE.
Furthermore, we also try to combine the TFD and DFD-D(w+u) together,
it's interesting to find that adding weakly and unlabeled data
doesn't bring any additional performance gain as we expected.

\subsubsection{Detail examination of the adaptive focal loss}
\label{subsubafl}

Table \ref{tab:result-afl} shows the detail examination results of
the proposed adaptive focal loss in section \ref{subsec:adapt-focal}.
The first line results represent the system 6 (PTS+TFD) results in
Table \ref{tab:result-all} without adaptive focal loss (single-stage
training, \verb"Stage 1"), and the setup with
$Epoch=15$ and $\gamma=2$ represents the system 8 (PTS+TFD+AFL) in
Table \ref{tab:result-all} that trained using two-stage training strategy
with the adaptive focal loss, this setup means that
in \verb"Stage 2", there are $200-15=185$ epochs are performed.
The $Epoch$ ($\tau$) is the first normal GL training
epochs that defined in \verb"Stage 1". The total training epoch is
fixed to 200 for all the systems.

\begin{table}[!htbp]
\caption{Development/Evaluation set $F1$-scores (\%) of the system training without (1st line, Stage 1) and with adaptive focal loss.}
\label{tab:result-afl}
\centering
\begin{tabular}{cc|ccc}
\toprule
Epoch ($\tau$) & $\gamma$ & Event-F1 & Segment-F1 & AT-F1   \\ \toprule
200   & - & 48.0~/~45.4        & 71.7~/~70.2          & 78.6~/~77.1 \\ \midrule
10    & 2 & \textbf{49.1}~/~47.0        & 75.4~/~\textbf{75.1}          & 78.0~/~76.6 \\
15    & 2 & 48.9~/~\textbf{47.1}        & \textbf{75.6}~/~74.0          & \textbf{80.0}~/~\textbf{78.1} \\
30    & 2 & 48.0~/~47.1        & 70.8~/~70.1          & 76.9~/~76.4 \\
15    & 1 & 45.5~/~44.8        & 74.3~/~73.3          & 75.3~/~74.4 \\
15    & 3 & 48.0~/~46.6        & 75.6~/~74.0          & 77.0~/~75.2 \\ \bottomrule
\end{tabular}
\end{table}

From Table \ref{tab:result-afl}, we see that only 15 epoches GL training
is enough to produce a relatively stable T-S model. During these epoches,
those easy acoustic events are already well-trained.
After this iteration, in \verb"Stage 2",  the model will be
guided to focus on the difficult-to-classify events by the introduced
adaptive focal loss in next training iterations. In addition,
it's clear that the focal loss is very sensitive to the scaling factor $\gamma$
of Equation (\ref{Laf}), because it controls the loss contribution
of each training clips for the total loss.
It is a very important parameter, too large or too small will make the model
produce different deviation. Therefore, the relatively best setup with
$\tau=15, \gamma=2$ is taken to train our final systems, such as system 8 and 9 in
Table \ref{tab:result-all}.
In addition, it can be seen from Table \ref{tab:result-afl} that the results
in the first row (single-stage training mode) of the performance comparison
of all systems using the two-stage training mode based on AFL are improved in
the segment based F1-score. At the same time, it can be seen that the impact
of entering the second stage in different epochs on the system is also very obvious.
Too early or too late will bring different performance.

\subsubsection{Validation of event-specific post processing}
\label{subsec:validesp}

\begin{figure*}[!thpb]
  \centering
  \scalebox{0.85}{\includegraphics[width=14cm]{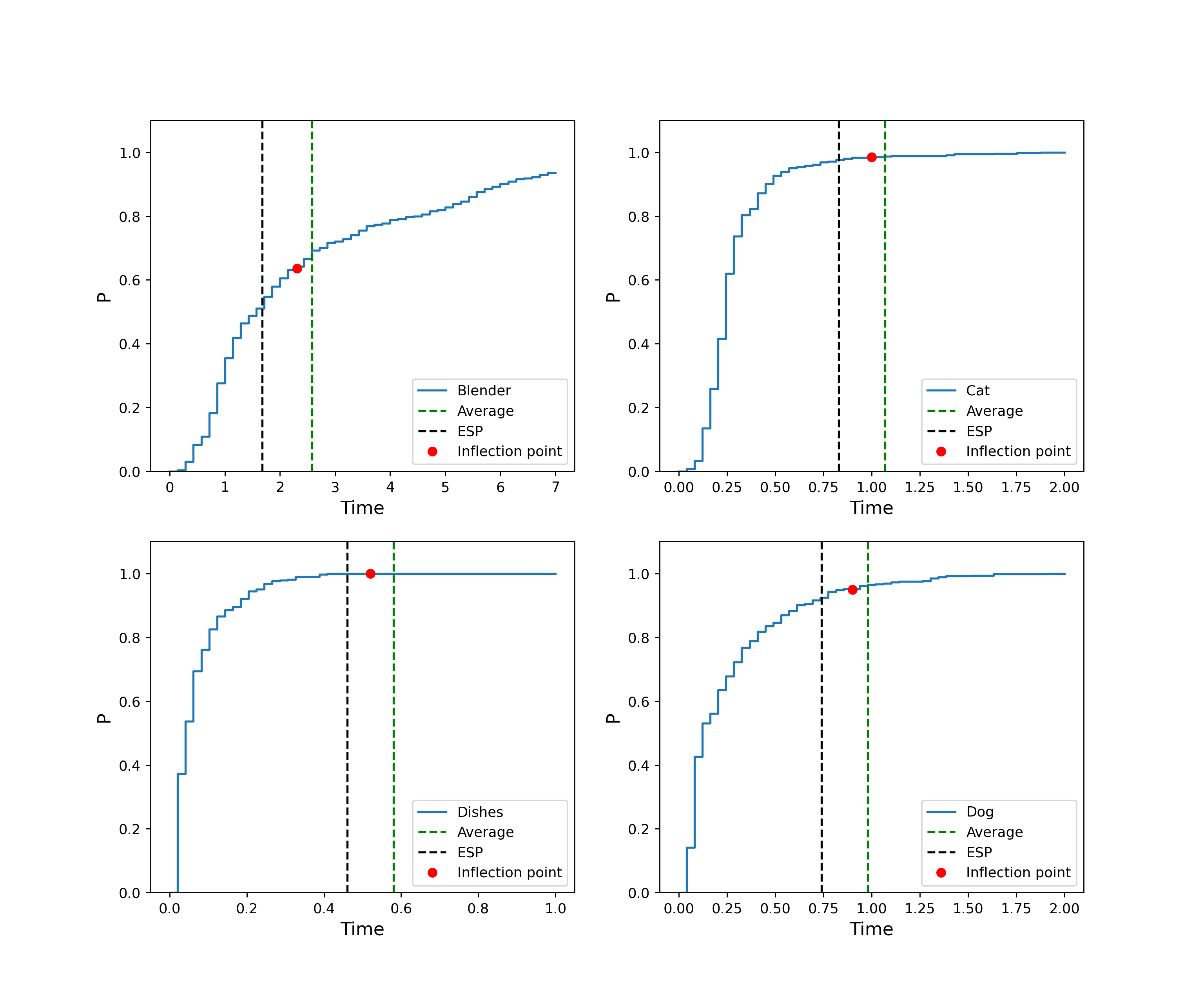}}
  \caption{Illustration of cumulative distribution of
  event-based duration that calculated from strong-labels for four different
  acoustic events: Blender, Cat, Dishes and Dog.}
  \label{fig:event-cdf}
\end{figure*}

Figure \ref{fig:event-cdf} illustrates four acoustic types' cumulative distributions of
event-based duration that calculated from strong-labels. The x-axis (abscissa axis) is the time-duration of each event, and the y-axis (ordinate axis) is the normalized occupation probability. As shown in Equation (7), the inflection point of this cumulative distribution is used to compute the ESP duration for each target event.
Comparing the four subfigures, we see that there is a big duration gap
for one event when it occurs in different clips. From each subfigure, we can see
the average duration (green line) of each event deviates far from the proposed ESP duration
(black line) that computed according to the event-based cumulative distributions in
Equation (\ref{eql-win}). Furthermore, from the four example distributions,
and indicated by the distribution changing of y-axis and the red inflection points,
it's clear that the durations of most clips
are more closer to the ESP durations than the average ones.
Meanwhile, the long flat-tails of these distributions vary with x-axis also indicates that,  there are obvious differences in the MF window size calculation by our proposed ESP and the traditional average way.
Therefore, we argue that the proposed ESP duration is better to reflect the
real distribution than the average one that used in most
previous works \cite{delphin2019mean,lin2020guided} to perform the median filtering
post processing.


\begin{table}[th]
\caption{Class-wise performances ($F1$-score(\%)) on the evaluation set with different post-processing methods. Bold fonts are used to highlight the $F1$-scores
with the largest performance improvements with respect to Avg and ESP.}
\label{tab:result-fl}
\centering
\begin{tabular}{@{}ccc@{}}
\toprule
Event class                & Avg(AED~/~AT) & ESP~(AED~/~AT) \\ \midrule
Alarm/bell/ringing         & 47.6~/~\textbf{77.4}   & 48.3~/~\textbf{83.3}   \\
Blender                    & 39.5~/~63.7   & 40.3~/~72.3   \\
Cat                        & \textbf{57.1}~/~86.1   & \textbf{66.2}~/~89.6   \\
Dishes                     & 34.5~/~74.6   & 38.5~/~78.2   \\
Dog                        & 52.2~/~87.7   & 52.8~/~88.3   \\
Electric shaver/toothbrush & 43.5~/~78.0   & 51.4~/~80.8   \\
Frying                     & 51.5~/~73.8   & 54.5~/~80.2   \\
Running water              & 32.9~/~68.1   & 31.2~/~63.2   \\
Speech                     & 56.9~/~93.4   & 58.9~/~93.3   \\
Vacuum cleaner             & 54.7~/~77.7   & 55.2~/~82.3   \\ \midrule
Overall                    & 47.1~/~78.0   & 49.8~/~81.2   \\ \bottomrule
\end{tabular}
\end{table}

Table \ref{tab:result-fl} demonstrates the class-wise performances with
different median filtering window size as the predictions post processing.
The `Avg' represents using adaptive median filtering window size with average duration
of each event class. `ESP' is our proposed event-specific post processing.
`AED/AT' is the event-based and class-wise $F1$-score for acoustic event detection
and audio tagging. It's clear to see that, the proposed ESP is very
effective to improve the system performances for most
target events, especially for those short-duration ones as
the blender, cat and dishes, etc. Such as, for the blender class,
the AT-F1 is significantly improved from 63.7\% to 72.3\%, and for the cat class,
the event-based AED $F1$-score is improved from 57.1\% to 66.2\%.
Compared with the `Avg' based median filtering, the proposed ESP improves the overall
AED/AT $F1$-score from 47.1/78.0\% to 49.8/81.2\%.

In addition, in Table \ref{tab:result-fl}, it is worth noting that,
on the real evaluation set, the class-wise F1-scores of all
the event classes except for running water have been improved.
This may due to the fact that, the MF window size of EPS
is calculated according to the strongly labeled data of 10 event classes,
however, the available strongly labeled data is very limited, and
there is heavy data imbalance between different classes.
For class `running water', the strongly labeled data amount is much less than that of
other classes. Therefore, the MF window size of EPS calculated for this class
may have a certain deviation from the actual length distribution
of the running water events,
which makes the proposed ESP for the running water prediction smoothing
fails to outperform the standard `Avg'.

\section{Conclusion}
\label{sec:conclution}

In this work, we propose a new joint framework for weakly supervised
acoustic event detection and audio tagging.  Different from previous
works that also based on the teacher-student learning
and CRNN frameworks, our proposal learns the models for
acoustic event detection and audio tagging tasks
as two independent branches in the whole architecture. Three new
methods are proposed to improve the joint framework, including
a frame-level target-events based deep feature distillation,
a two-stage training strategy with adaptive focal loss, and
an event-specific post processing. All of these techniques are validated
on the dataset of DCASE 2019 Task 4 challenge, experimental results
show that the new joint framework with all the proposed techniques
achieves competitive performances in both acoustic event detection
and audio tagging tasks.
On the real evaluation set of DCASE 2019 Task 4 challenge,
our proposed system significantly outperforms the best single submission
system (GL) by absolute 9.3\%, 9.4\% and 11.0\% F1-scores in event-based,
segment-based AED and macro-average AT-F1, respectively.
Our future work will focus on generalizing the proposed framework to other acoustic event
detection tasks, try our separate models for AT and AED tasks with all the proposed
techniques, and improving the event-specific post processing
to fix the strongly labeled data imbalance issue.

\section*{Acknowledgment}

This work was funded by the National Natural Science Foundation of China (Grant No.62071302).

\bibliography{mybibfile}

\end{document}